\documentclass[aps,pre,showpacs,showkeys,amssymb]{revtex4-2}

\usepackage{bm}
\usepackage{graphicx}
\usepackage{amsfonts}
\usepackage{amsmath}
\usepackage{xcolor}
\usepackage{verbatim}

\begin{document}

\title{Simplified energy landscape of the $\phi^4$ model and the phase transition}

\author{Fabrizio Baroni}
\email{f.baroni@ifac.cnr.it, baronifab@libero.it}
\affiliation{IFAC-CNR Nello Carrara Institute of Applied Physics, Sesto Fiorentino (FI), Italy}

\date{\today}

\begin{abstract}
The on-lattice $\phi^4$ model is a paradigmatic example of a continuous real-variable model undergoing a continuous symmetry braking phase transition (SBPT). Here, we study the $\mathbb{Z}_2$-symmetric mean field case without a quadratic term in the local potential. We show that the $\mathbb{Z}_2$-SBPT is not affected by the quadratic term and that the potential energy landscape is greatly simplified from a topological viewpoint. In particular, only three critical points exist to confront with a number growing as $e^N$ ($N$ is the number of degrees of freedom) of the model with negative quadratic term. 
This may be a crucial feature of the $\phi^4$ model because in recent years the study of the link between statistical mechanic and geometric-topological properties of configuration space has received an increasing attention. 
In particular, we focus on the properties of the equipotential hypersurfaces with the aim of deepening the link between the SBPTs and the essential properties of the potential landscape capable to entail them. 
\end{abstract}



\maketitle


\section{Introduction}
\label{intro}

This study is part of a research line attempting to clarify the relationship between phase transitions and the potential energy landscapes of Hamiltonian systems (for example, see \cite{cpc,ccp1,fcsp,ck,b1,b6,rs,ckn,dhk,fp,kss}). Potential landscapes include critical points, geometric properties, and topology of suitable subsets of configuration space, for example, equipotential hypersurfaces. This type of study often makes intensive use of models undergoing phase transitions, including the $\phi^4$ model. This model has received an increasing attention in the last years, and is a paradigmatic example of a model undergoing a continuous phase transition (for example, see \cite{aarz,bc,b0,b3,garanin,hk1,hk,mhk,km,k1,gfp2}). Here, we introduce a simplified version of the model that shows a reduced number of critical points compared with the traditional version. 

The $\phi^4$ model is a lattice version of a classical $\phi^4$ field model. This can be studied in any spatial dimension, in scalar and vector versions by the Hamiltonian
\begin{eqnarray}
   H=\sum_{\alpha=1}^n\sum_{i=1}^N\left[\frac{1}{2}\left(\pi^{\alpha}_i\right)^2-\frac{\mu}{2}\left(\phi^{\alpha}_i\right)^2-J\sum_{\left\langle i,j\right\rangle} \phi^{\alpha}_i\phi^{\alpha}_{j}\right]\nonumber
	+\frac{\lambda}{4}\sum_{i=1}^N\left[\sum_{\alpha=1}^n\left(\phi^{\alpha}_i\right)^2\right]^2,
	\label{phi4model}
\end{eqnarray}
where the index $\alpha$ runs from $1$ to $n$ for an $O(n)$ symmetry group, the index $i$ labels the $d$-dimensional spatial lattice, $(\pi_i, \phi_i)$ are the canonically conjugated variables, $N$ is the number of degree of freedom, and $\left\langle i,j\right\rangle$ is the set of the nearest-neighbor lattice sites of the $ith$ site \cite{pettini}. The set $\left\langle i,j\right\rangle$ can be defined in other ways, for example, the set of all the variables within a certain range, or the whole lattice in the case of mean-field interactions. 

The model is known to undergo an $O(n)$-symmetry breaking phase transition (SBPT). The existence of a SPBT can be proven using renormalization group arguments \cite{wk}. For $d=2$ and $n=2$ and according to the Mermin-Wagner theorem, the model cannot have any SBPT because of the combination of short-range interactions, continuous symmetry, and two spatial dimensions. In fact, it undergoes a Kosterlitz-Thouless phase transition without affecting the order parameter which remains vanishing even below the critical temperature.

In \cite{b0}, the topology of the equipotential hypersurfaces interior of the mean-field $\phi^4$ model of an $O(1)$ symmetry (also called $\mathbb{Z}_2$) was solved using Morse theory. A large number of critical points that exponentially increase with $N$ were identified. In \cite{km, dhk}, a similar study was conducted using the nearest-neighbor-$2$d version. At sufficiently small values of the coupling $J$ there is no difference in the number of critical points compared to the mean-field version, but by increasing $J$ and leaving fixed $N$, their number rapidly drops to three only.

We founded the negative quadratic term of the local potential of the mean-field $\phi^4$ model to be responsible of the rapid growth of the critical points number while increasing $N$. Knowing this, we ask if the presence of that term can be justified in order to entail the SBPT. Firstly, it derives from classical field theory, where it is the mass term of the associated classical field, and for this reason it is often labeled as $\mu^2>0$. Another reason to justify the presence of the negative quadratic term is the wish to simulate the classical spin of the Ising model. Indeed, for $n=1$ the $\phi^4$ model can be seen as a continuous-variable version of the classical Ising model whose classical spins, $S_i$'s, can take two values only: generically assumed as $\pm 1$. In the $\phi^4$ model the role of the two permitted values of the $S_i$'s are played by the two global minima of the double-well local potential $V_l(\phi)=\lambda\phi^4-\mu\phi^2$ entailed by the negative quadratic term.  

Anyway, there is a remarkable difference between the $S_i$'s and the $\phi_i$'s: the former do not find any resistance at jumping between the two permitted values, while the latter find such a resistance at jumping between the two global minima because of the presence of the potential barrier. Hence, in our opinion, the double well introduces a complication which is not present in the classical Ising model, and furthermore it is not even necessary for entailing the SBPT. For these reasons, in the following we allow $\mu$ to be vanishing or, in some particular conditions, negative.

In Sec. \ref{pphi4mf} we set $\mu$ to zero in the $\phi^4$ model and we study the mean-field version with a $\mathbb{Z}_2$ symmetry because it is the simplest case with both canonical thermodynamic and critical points of configuration space solvable in a semi-analytical way. In Sec. \ref{phi4noint} we study the same model without interacting potential where no SBPT occurs in order to make a comparison. In Sec.  \ref{short} we consider some short-range versions and find out all their critical points using the NPHC method up to $N=9$.

\section{Mean-field $\phi^4$ model with a vanishing quadratic term in the local potential}
\label{pphi4mf}

In what follows we disregard the kinetic terms $\pi^2_i/2$, $i=1,\cdots,N$, in the Hamiltonian (\ref{phi4model}) because they yield a trivial contribution to the partition function which can be factorized, and from a topological viewpoint in phase space the level sets at constant kinetic energy are trivially equivalent to $N$-spheres. Then we set the parameters $\lambda=2/N$, $\mu=0$ and extend the interaction to all the pairs of coordinates (i.e., mean-field interaction), giving rise to the potential
\begin{equation}
V=\frac{1}{4}\sum_{i=1}^N \phi_i^4-\frac{J}{2N}\left(\sum_{i=1}^N \phi_i\right)^2.
\label{Vmf}
\end{equation}

\subsection{Canonical thermodynamic}
\label{pphi4mf_thermo}

\begin{figure}
\begin{center}
\includegraphics[width=0.35\textwidth]{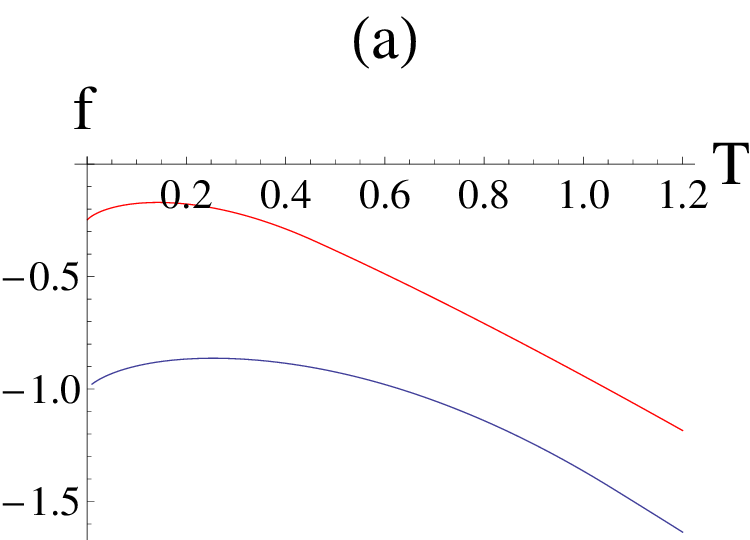}
\includegraphics[width=0.35\textwidth]{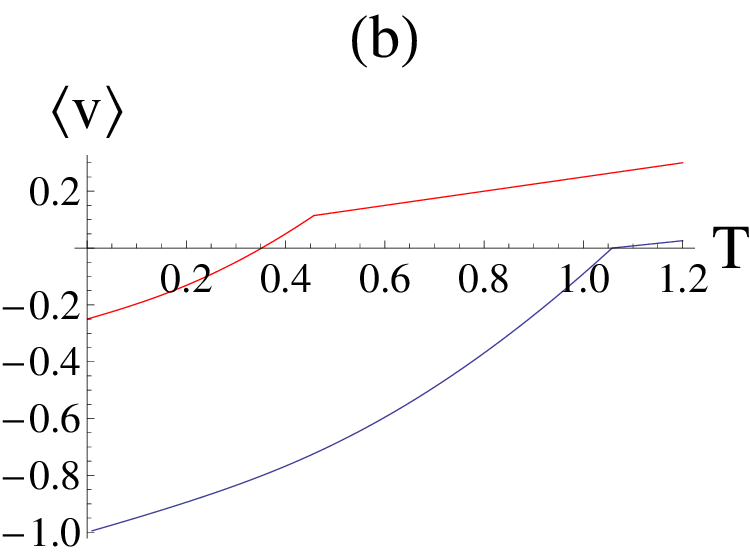}
\includegraphics[width=0.35\textwidth]{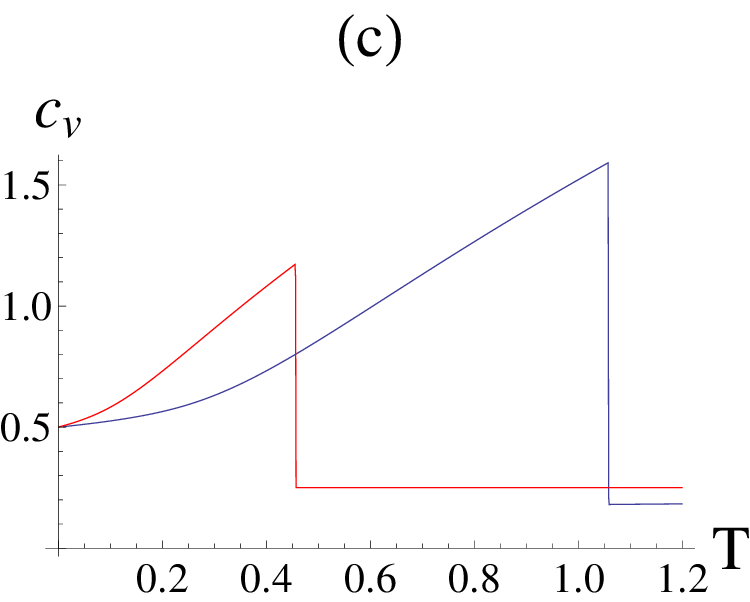}
\includegraphics[width=0.35\textwidth]{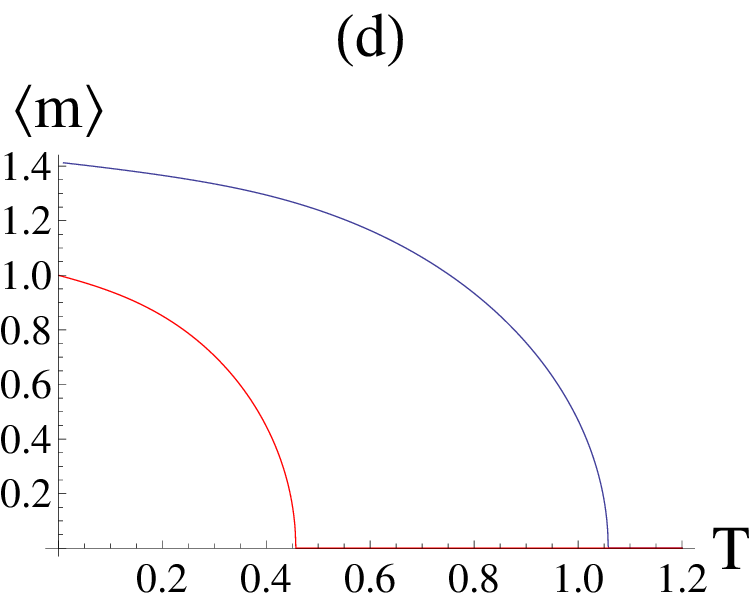}
\caption{(a), (b), (c) and (d) are, respectively, free energy, specific average potential, configurational specific heat and spontaneous magnetization as functions of the temperature. The red lines are for the model (\ref{Vmf}) and the blue lines for the model (\ref{Vtradmf}), both with coupling constant $J=1$.}
\label{pphi4_thermo_fig}
\end{center}
\end{figure}

\begin{figure}
\begin{center}
\includegraphics[width=0.35\textwidth]{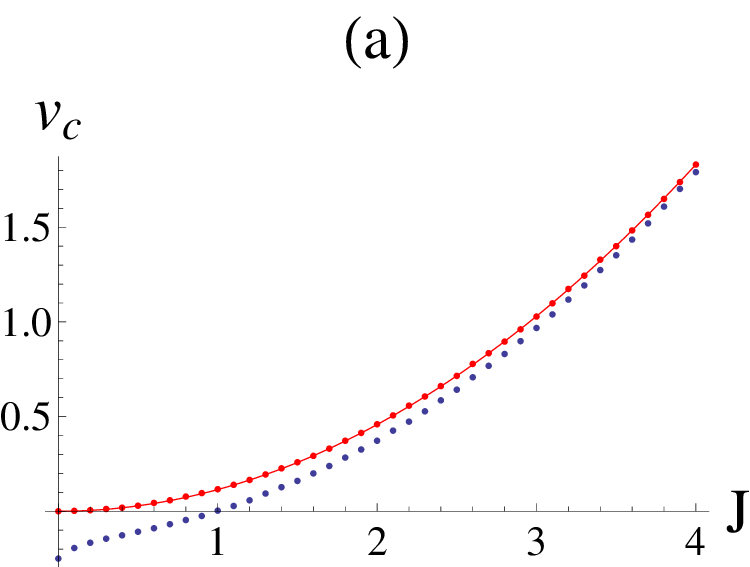}
\includegraphics[width=0.35\textwidth]{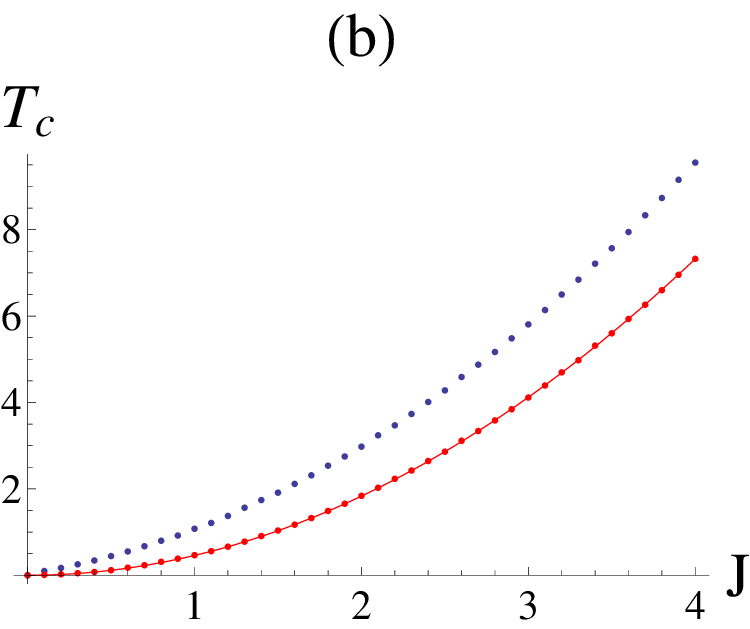}
\caption{(a) Thermodynamic critical average potential as a function of the coupling constant $J$ of model (\ref{Vmf}) (red points) and of model (\ref{Vtradmf}) (blues points). The continuous line is the parabola $0.114446J^2$ fitted to the data. (b) As (a) for the critical temperature. The continuous line is the parabola $0.457786J^2$.}
\label{pphi4_TcJ_fig}
\end{center}
\end{figure}

In \cite{b0,dl}, the thermodynamic of the model (\ref{phi4model}) with $\lambda=2/N$, $\mu=1$, and mean-field interactions, i.e. with Hamiltonian
\begin{equation}
V=\sum_{i=1}^N \left(\frac{1}{4}\phi_i^4-\frac{1}{2}\phi_i^2\right)-\frac{J}{2N}\left(\sum_{i=1}^N \phi_i\right)^2,
\label{Vtradmf}
\end{equation}
was solved using the mean-field theory. In this section we will follow the same path way for the model (\ref{Vmf}). In Figs. \ref{pphi4_thermo_fig} and \ref{pphi4_TcJ_fig} the results for these two models are put in comparison. 

The configurational partition function is 
\begin{equation}
Z=\int\,d^N\phi\,e^{-\beta\left[\sum_{i=1}^{N}V_{loc}(\phi_i)-\frac{J}{2N}\left(\sum_{i=1}^N \phi_i\right)^2\right]},
\end{equation}
where
\begin{equation}
V_{loc}(\phi)=\frac{\phi^4}{4}
\label{vlocal}
\end{equation}
is the local potential. The order parameter, i.e. the magnetization in our case, is
\begin{equation}
m=\frac{1}{N}\sum_{i=1}^N\phi_i,
\label{m}
\end{equation}
which, replaced in $Z_c$, gives
\begin{equation}
Z=\int d^N\phi\,e^{-\beta\left[\sum_{i=1}^{N}V_{loc}(\phi_i)-\frac{JN}{2} m^2\right]}.
\label{Zc}
\end{equation}
The fact that the mean-field interactions imply the interacting potential is a function of the magnetization, allows us to analytically solve $Z$ using the Hubbard-Stratonovich transformation \cite{goldenfeld} based on the equality

\begin{equation}
    e^{\mu m^2}=\frac{1}{\sqrt{\pi}}\int dy\,e^{-y^2+2\sqrt{\mu}my},
\end{equation}
which, inserted in (\ref{Zc}), yields
\begin{equation}
    Z=\frac{1}{\sqrt{\pi}}\int dy\left[\int d\phi\,e^{-\beta V_{loc}(\phi)+\sqrt{\frac{2\beta J}{N}}m\phi}\right]^N e^{-y^2}.
\end{equation}
After introducing
\begin{equation}
    \varphi(m,\beta)=\ln\int dq\,e^{-\beta \left[V_{loc}(q)+J m q\right]},
\end{equation}
and the variable changing $y=\sqrt{\frac{N\beta J}{2}}m$, we get
\begin{equation}
    Z=\sqrt{\frac{N\beta J}{2\pi}}\int dm\,e^{-N\beta f(m,\beta)},
\end{equation}
where
\begin{equation}
    f=-\frac{J}{2}m^2+T\varphi (m,T)
\end{equation}
is the configurational Helmholtz free energy per degree of freedom. 

Finally, in order to apply the saddle point method to calculate $Z$, we minimize $f$ with respect to $m$ at fixed $T$ obtaining the spontaneous magnetization $\left\langle m\right\rangle(T)$. From the latter, we get the free energy, the average potential, and the specific heat
\begin{eqnarray}
    f(T)&=&-\frac{1}{N\beta}\ln Z,
		\\
		\left\langle v\right\rangle(T)&=&-\frac{\partial}{\partial\beta}Z,
		\\
		c_v(T)&=&\frac{d\left\langle v\right\rangle}{dT},
\end{eqnarray}
respectively. They are plotted in Fig. \ref{pphi4_thermo_fig} in comparison with the results for the mean-field $\phi^4$ model (\ref{Vtradmf}). The picture is the well known one of a second-order $\mathbb{Z}_2$-SBPT with classical critical exponents. 

We cannot see any difference in the thermodynamic of the two models, apart from a quantitative viewpoint. We conclude that the negative quadratic term in the local potential has no part in causing the SBPT. This is not surprising because in \cite{b1}, it was showed that in a mean-field model a double-well potential with a minimum barrier between the wells proportional to $N$ is a sufficient condition for entailing a SBPT. Both the models (\ref{Vmf},\ref{Vtradmf}) have this feature independently of the presence of the quadratic term $-\phi^2/2$. Rather, as we will see in the following, the latter yields complication and confusion about the real connection between the characteristic of the equipotential hypersurfaces of configuration space and the SBPT. The occurrence of the double well of the potential is generated by the competition between the confining part given by the local potential and the interacting part for $J>0$. The only essential condition to satisfy in order to make the total potential confining is the following 
\begin{equation}
\lim_{\phi\rightarrow+\infty}\frac{V_{loc}(\phi)}{\phi^2}=+\infty.
\end{equation}

For example, let us consider the local potential given by the square well
\begin{equation}
V_{loc}=\begin{cases}
+\infty& \text{ if }\quad |\phi|\geq 1 
\\ 
0& \text{ if }\quad |\phi|<1
\end{cases},
\end{equation}
which is nothing but the limit of $V_{loc}=\phi^{2k}$ for $k\rightarrow\infty$ with $k$ natural. By this choice, we get in the thermodynamic limit the configurational partition function of the mean-field Ising model, whose free energy is given by
\begin{equation}
  f(m,T)=-\frac{J}{2}m^2+1+T\ln\cosh\left(\frac{J m}{T}\right).
\end{equation}
From the latter, by setting to zero the derivative with respect to $m$, we get the spontaneous magnetization as the solutions of the following equation
\begin{equation}
  -m+\tanh\left(\frac{J m}{T}\right)=0.
\end{equation}
The critical temperature is $T_c=J$.

\begin{figure}
\begin{center}
\includegraphics[width=0.35\textwidth]{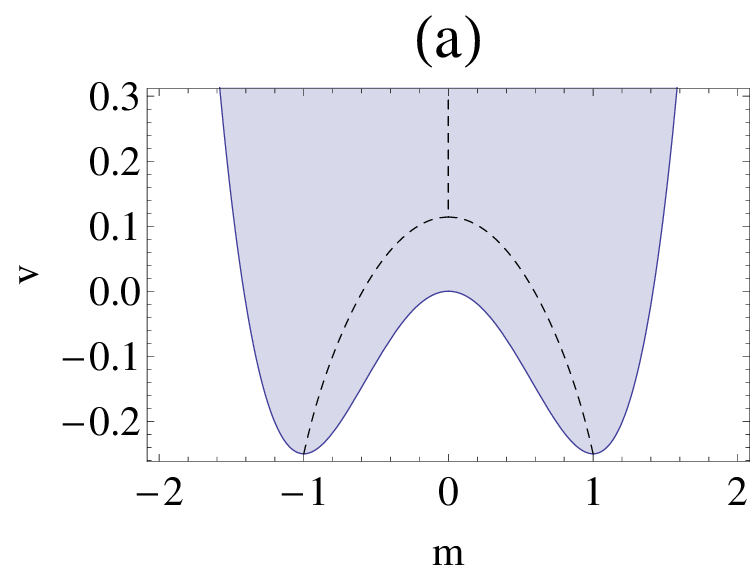}
\includegraphics[width=0.35\textwidth]{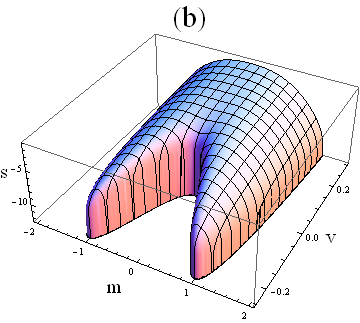}
\caption{(a) Domain of the entropy $s$ of the model (\ref{Vmf}) (dark region) in the $(m,v)$-plane. The dashed line is the spontaneous magnetization where $s$ takes the maximum evaluated along straight lines with constant $v$. (b) $3$D plot only in qualitative accordance for what concerns the s-values with the graph (a).}
\label{pphi4_sdomain_fig}
\end{center}
\end{figure}

In \cite{hk1}, the large deviation theory was applied to find out the entropy $s(v,m)$ of the mean-field $\phi^4$ model (\ref{Vtradmf}) and of the same model without interaction. That theory can be applied to the model (\ref{Vmf}) as well. We expect no qualitative difference in the properties of $s(v,m)$ which is analytic and non-concave. The non-concavity is allowed by the long-range interactions and is strictly related to the SBPT, whereas it is forbidden in the short-range case \cite{lanford,galla,hk,k2}. In simple terms, an equilibrium configuration of a short-range system in the broken symmetry phase can be divided by a layer into two domains each with magnetization oriented independently of the other. This is made possible because the potential energy at the layer becomes negligible compared with the total one in the thermodynamic limit. Here, we limit to give the domain of $s(v,m)$, whose contour is given by the potential evaluated on the straight line in configuration space passing by the origin of coordinates and orthogonal to the hyperplanes at constant $m$: $v(m)=m^4/4-Jm^2/2$. It is plotted in Fig. \ref{pphi4_sdomain_fig}.

\subsection{Critical points and topology of the equipotential hypersurfaces}
\label{pphi4mf_geom}

$\nabla V=0$ for the potential (\ref{Vmf}) takes the form
\begin{equation}
   \phi^3_i-\frac{J}{N}\sum^{N}_{i=1}\phi_i=0\quad\quad i=1,\cdots,N.
	\label{gradmf}
\end{equation}
The form of the system (\ref{gradmf}) implies that the components of the solutions are all equal, so that it reduces to $\phi^3_i-J\phi_i=0$, $i=1,\cdots,N$. Trivially, the solutions are $\phi_0^s=(0,\cdots,0)$, and $\phi_{\pm}^s=\pm\sqrt{J}\left(1,\cdots,1\right)$.

The equipotential hypersurfaces are often called in literature $\Sigma_{v,N}$'s and are defined as follows
\begin{equation}
	\Sigma_{v,N}=\{\mathbb{\phi}\in \mathbb{R}^N: \frac{V(\phi)}{N}=v\}.
	\label{ES}
\end{equation}
In order to apply Morse theory, it is useful to define also $M_{v,N}$ as
\begin{equation}
	M_{v,N}=\{\mathbb{\phi}\in \mathbb{R}^N: \frac{V(\phi)}{N}\leq v\}.
	\label{M}
\end{equation}
Trivially, the $\Sigma_{v,N}$'s are the boundary of the $M_{v,N}$'s:
\begin{equation}
	\Sigma_{v,N}=\partial M_{v,N}.
\end{equation}
The topology of the $\Sigma_{v,N}$'s is strictly related to that of the $M_{v,N}$'s.

According to Morse theory, the topology of the $M_{v,N}$'s can be determined starting from a Morse function defined on configuration space. A Morse function is a function whose critical points are non-degenerate, i.e. isolated. In this case, we use the potential as a Morse function. Once the critical points have been found, the topology of the $M_{v,N}$'s is retrieved attaching a $k$-handle $\texttt{H}^{N,k}$ for each critical point, where $N$ is the configuration space dimension and $k$ is the index of the critical point ($0\leq k\leq N$). $\texttt{H}^{N,k}$ is the product of two disks, one $k$-dimensional and the other $(N-k)$-dimensional
 \begin{equation}
	\texttt{H}^{N,k}=D^k\times D^{N-k}.         
\end{equation}
The index of a critical point is defined as the number of negative eigenvalues of the Hessian matrix $H$, which for the potential (\ref{Vmf}) takes the form
 \begin{equation}
   H_{ij}=\frac{\partial^2 V}{\partial\phi_i\partial\phi_j}=3\phi_i^2\delta_{ij}-\frac{J}{N}.         
\end{equation}
The fact that the critical points are non-degenerate are equivalent to request that each critical point is non-singular, or in other words, that the determinant of the Hessian matrix is non-vanishing. For $\phi_{\pm}^s$, $H_{ij}=3J\delta_{ij}-J/N$, which entails the index is $0$ because all the eigenvalues are positive. For $\phi_0^s$, $H_{ij}=-J/N$, which entails the saddle is singular. Anyway, the saddle is isolated and it can be showed that it corresponds to a critical point with index $1$. Consider an orthonormal coordinate system such that an axis is the line passing through the points $\phi_{\pm}^s$ and the remaining axes are orthogonal to the latter. The second derivative of $V$ along the aforementioned axis computed at $\phi_0^s$ is negative. The restriction of $V$ on each of the other axes has a global minimum in $\phi_0^s$. From this two consideration, we can infer that $\phi_0^s$ corresponds to a critical point with index $1$. 

\begin{figure}
\begin{center}
\includegraphics[width=0.35\textwidth]{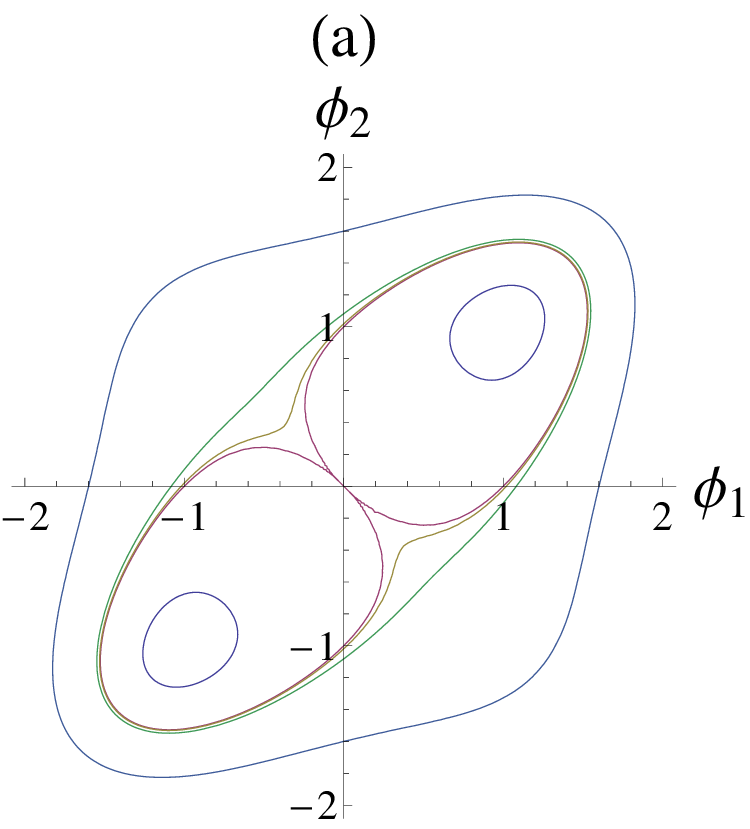}
\includegraphics[width=0.35\textwidth]{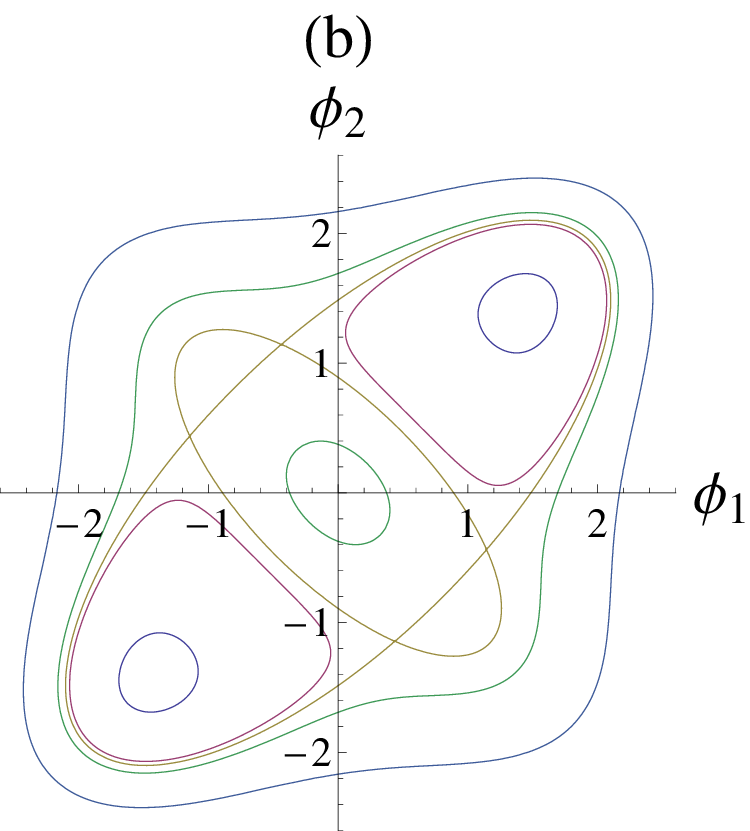}
\caption{Some $\Sigma_{v,N}$'s of the model (\ref{Vmf}) at $v=-0.4, 0, 0.01, 0.05, 0.2$ for $N=2$ with $J=1$ (panel (a)) in comparison with the ones at $v=-1, -0.495,-0.4375, -0.2, 1$ of the model (\ref{Vtradmf}) (panel (b)). The proliferation of the critical points in the latter is already evident at $N=2$. In panel (a), $\Sigma_{0.05,2}$ is the precursor of the $\Sigma_{v,N}$ which, at any $N$, is the boundary between the 'dumbbell-shaped' ones and those which are not.}
\label{sigmav}
\end{center}
\end{figure}

The critical points $\phi_{\pm}^s$ correspond to the global minimum of the potential $v_{min}=-J^2/4$ to which the first critical level starting from bottom corresponds. The critical point $\phi_0^s$ is a saddle point and corresponds to the $0$-critical level, i.e. the second and last critical level from top to bottom to top.
The topology of the $M_{v,N}$'s are retrieved attaching two $0$-handles, $\texttt{H}^{N,0}$, at the first critical level and an $1$-handle, $\texttt{H}^{N,1}$, at the second critical level. Therefore, the $M_{v,N}$'s are homeomorphic to a couple of disjointed $N$-balls for $v\in (-J^2/4,0)$, whereas for $v\in (0,+\infty)$ they are homeomorphic to a single $N$-ball. Equivalently, the topology of the $\Sigma_{v,N}$'s is that of two $N$-spheres for $v\in (-J^2/4,0)$ and to an $N$-sphere for $v\in (0,+\infty)$ (see Fig. \ref{sigmav}).

\begin{figure}
	\begin{center}
		\includegraphics[width=0.647\textwidth]{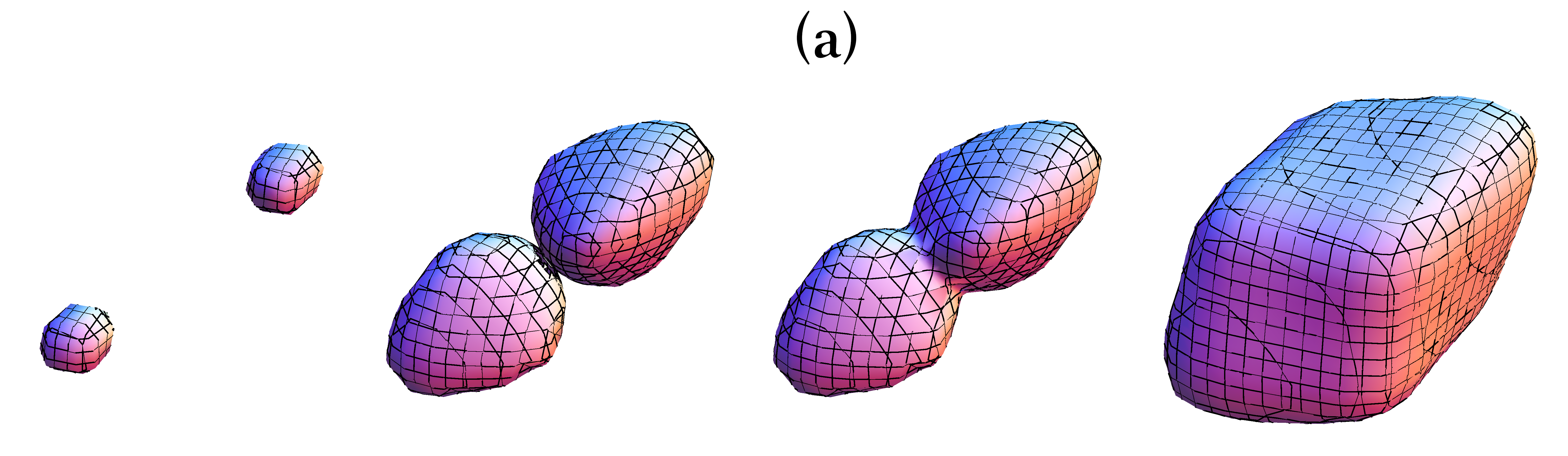}
		\includegraphics[width=0.647\textwidth]{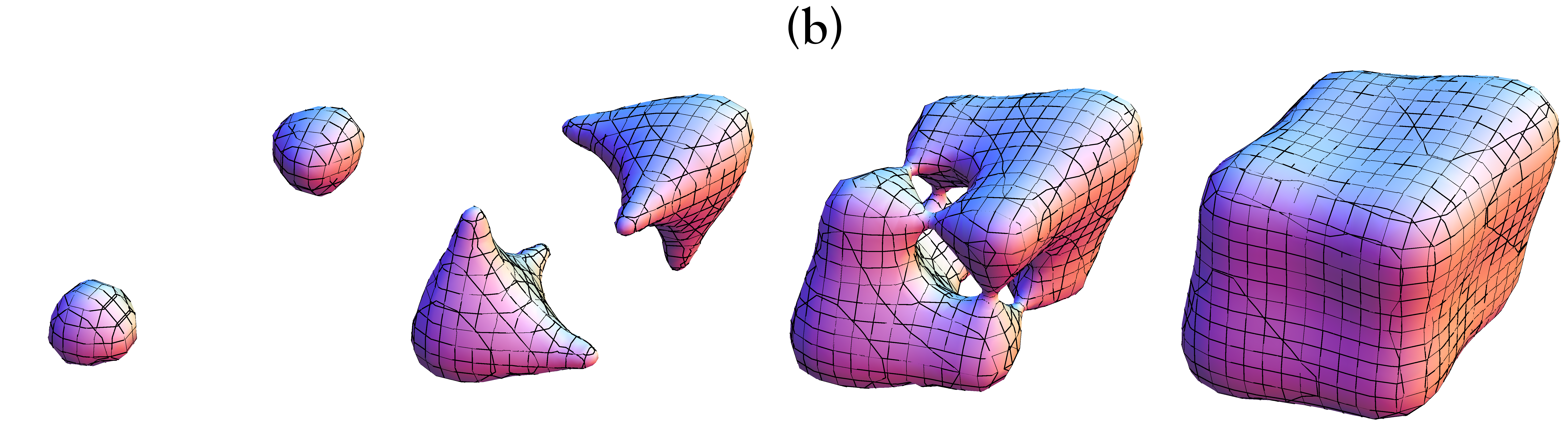}
		\caption{(a) Some $\Sigma_{v,N}$'s for the model (\ref{Vmf}) for $N=3$ and $J=1$. The potential increases from left to right. (b) The same of panel (a) for the model (\ref{Vtradmf}).}
		\label{sphi4ES}
	\end{center}
\end{figure}

\begin{figure}
\begin{center}
\includegraphics[width=0.35\textwidth]{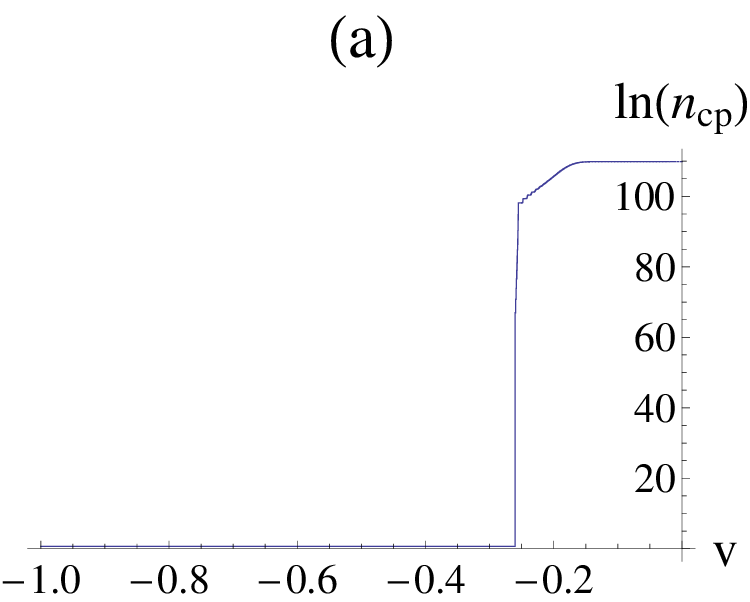}
\includegraphics[width=0.35\textwidth]{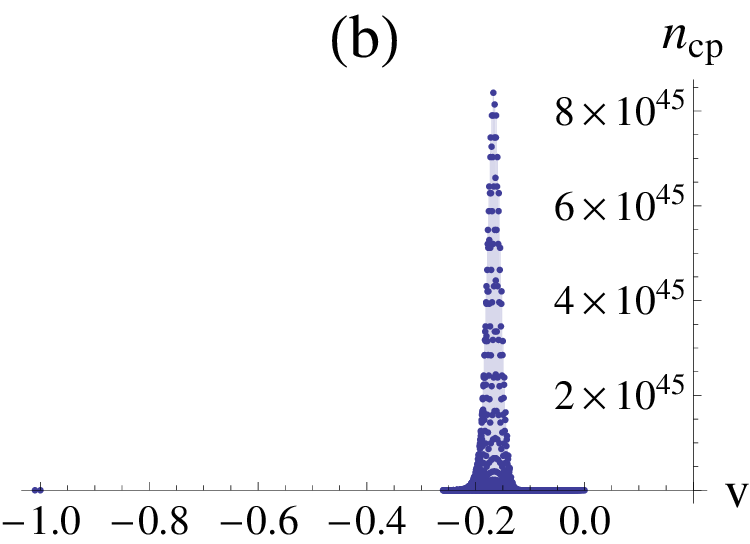}
\caption{Model (\ref{Vtradmf}) for $N=100$ and $J=1$. (a) Logarithmic sum of critical points starting from left versus potential density. (b) Number of critical points with respect to their critical potential density.}
\label{pcphi4}
\end{center}
\end{figure}

In \cite{b0}, the critical points of the model (\ref{Vtradmf}) were found using a semi-analytic method. Fig. \ref{pcphi4} shows the number of critical points and their density with respect to the potential density. The total number is of the order of $e^N$, at least for the $J$-values investigated always less or equal to $1$. We cannot exclude that by increasing $J$ with fixed $N$ a decreasing of the number of critical points can occur similarly to what was detected in \cite{dhk} for the short range case. The critical points are comprised between a minimum $v$-value, significantly greater than the global minimum, and $v=0$. In \cite{b0}, it was also given an analytic demonstration that there is no critical point above $v=0$. From a topological viewpoint of the $\Sigma_{v,N}$'s, the interval of $v$-values containing the critical points works as a transition between two $N$-spheres and a single $N$-sphere (see Fig. \ref{sigmav} and \ref{sphi4ES}). In the model (\ref{Vmf}) this transition interval collapses in a single critical level with a single critical point. It is notable that, despite this dramatic simplification of the potential landscape, the model (\ref{Vmf}) does not lose any property from a phase transition viewpoint.



\begin{figure}
	\begin{center}
		\includegraphics[width=0.35\textwidth]{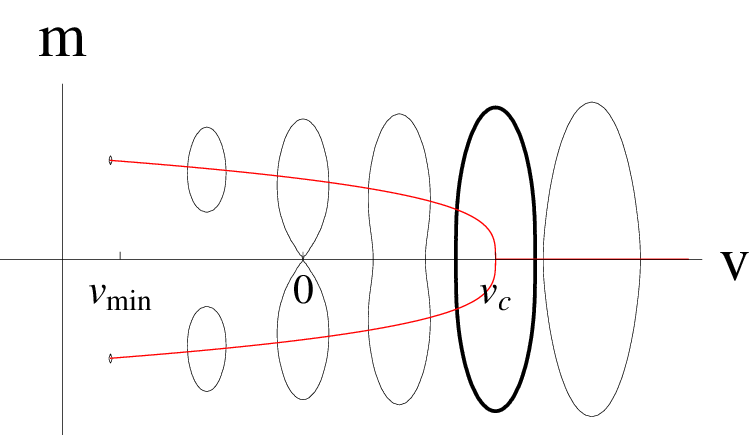}
		\caption{Picture of the symmetry-breaking mechanism based on 'dumbbell-shaped' $\Sigma_{v,N}$'s for the model (\ref{Vmf}). The dumbbell-shaped $\Sigma_{v,N}$'s are below the thermodynamic critical potential $v_c$. In red the spontaneous magnetization.}
		\label{magsigmav}
	\end{center}
\end{figure}

In \cite{b3,b4}, it was showed that a $\mathbb{Z}_2$-SBPT can be entailed by 'dumbbell-shaped' $\Sigma_{v,N}$'s. Ruoghly speaking, such a $\Sigma_{v,N}$ is made up of two major lobes connected by a narrow neck (see Fig. \ref{sphi4ES} and \ref{magsigmav}). In more details, a $\Sigma_{v,N}$ is dumbbell-shaped when the microcanonical volume of the section at constant $m$ does not take the global maximum at $m=0$. The critical potential corresponds to the transition between the dumbbell-shaped $\Sigma_{v,N}$'s and those witch are not. This $\mathbb{Z}_2$-SBPT generating-mechanism was discovered in several models \cite{bc,b1,b2,b3,b4,b6} and it is acting also in the model (\ref{Vmf}) and (\ref{Vtradmf}). The picture is represented in Fig. \ref{magsigmav}. The method used in Sec. \ref{pphi4mf_thermo} to solve the canonical thermodynamic decomposes the $N$-dimensional integral of the partition function in an integral with r

pect to $m$. In this way, the problem to find the microcanonical volume of the sections of the $\Sigma_{v,N}$'s at constant $m$ was solved.  

We remark that being dumbbell-shaped for a $\Sigma_{v,N}$ can be independent at all on critical points and topology. Anyway, there is a special case that it is worth  mentioning, i.e. when the $\Sigma_{v,N}$ is made up from two or more connected components which do not intersect the hyperplane at $m=0$. In this case, the $\Sigma_{v,N}$ is necessarily dumbbell-shaped for obvious reasons. This is the case of the model (\ref{Vmf}) for $v$ comprised between the global minimum and $0$ and of the model (\ref{Vtradmf}) for $v$ comprised between the global minimum and the minimum $v$-value of the critical $\Sigma_{v,N}$'s. In fact, the $\Sigma_{v,N}$'s are all homeomorphic to two $N$-spheres. The fact that a dumbbell-shaped $\Sigma_{v,N}$ impli

 the $\mathbb{Z}_2$ symmetry breaking has a remarkable consequence on the critical potential. In particular, it has to be greater than $0$ for the model (\ref{Vmf}) and greater than the minimum of the $v$-values of the critical $\Sigma_{v,N}$'s for the model (\ref{Vtradmf}). Both the inferences are compatible with the results obtained here and in \cite{b0}.

$v_c>0$ is a consequence of Theorem 1 in \cite{bc} which states that: if the $\Sigma_{v,N}$'s are made up of two, or more, disjointed connected components non-intersecting the hyperplane at $m=0$ below a certain value $v_0$ of the potential density, then the symmetry $\mathbb{Z}_2$ is broken for any $v<v_0$. In the case of the model (\ref{Vmf}), $v_0=0$. This condition is a special case of the sufficient condition given in Theorem 1 in \cite{b3} based on dumbbell-shaped $\Sigma_{v,N}$'s. This because if the $\Sigma_{v,N}$'s are made up as just described, the more reason they are dumbbell-shaped.

\section{Model (\ref{Vmf}) without interaction}
\label{phi4noint}

To make a comparison with a model without SBPT, the interacting terms were removed from the Hamiltonian (\ref{Vmf})
\begin{equation}
	V=\frac{1}{4}\sum_{i=1}^N\phi_i^4.
	\label{Vnoint}
\end{equation}
The solution of the canonical thermodynamic is trivial because the system is nothing but a collection of $N$ independent quartic oscillators. No phase transition can occur. The partition function is given by
\begin{equation}
Z_c=\left(\int d\phi\,e^{-\frac{1}{4}\beta N\phi^4}\right)^N=\left(\frac{\gamma\left(\frac{1}{4}\right)}{\sqrt{2}}T^\frac{1}{4}\right)^N,
\end{equation}
from which we get the caloric curve $\left\langle v\right\rangle(T)=T/4$.

The topology of configuration space is even more trivial, indeed, $\nabla V=0$ takes the simple form
\begin{equation}
   \phi^3_i=0\quad\quad i=1,\cdots,N,
\end{equation}
whose unique solution is $(0,\cdots,0)$. The index cannot be computed as for a Morse function because the Hessian matrix vanishes at $(0,\cdots,0)$. However, since $(0,\cdots,0)$ is a global minimum, it corresponds to a non-singular stationary point with index $0$. From a topological viewpoint, the topology of the $M_{v,N}$'s for $v>0$ can be retrieved attaching a $0$-handle, $\texttt{H}^{0,N}$, at the $0$-critical level. Hence, the $M_{v,N}$'s are homeomorphic to an $N$-ball for any $v>0$. The model cannot undergo to any SBPT, not even at $T=0$ because at that temperature the representative point is frozen at $(0,\cdots,0)$ to which a vanishing spontaneous magnetization corresponds. 

\begin{figure}
\begin{center}
\includegraphics[width=0.35\textwidth]{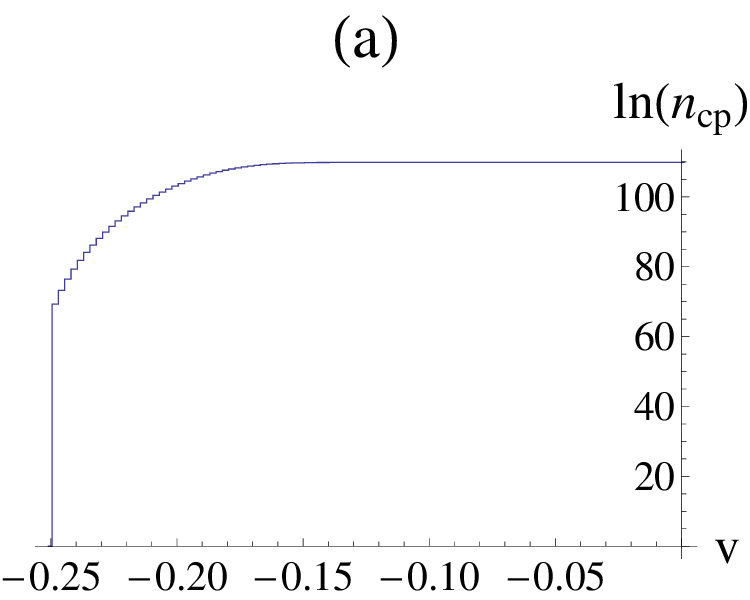}
\includegraphics[width=0.35\textwidth]{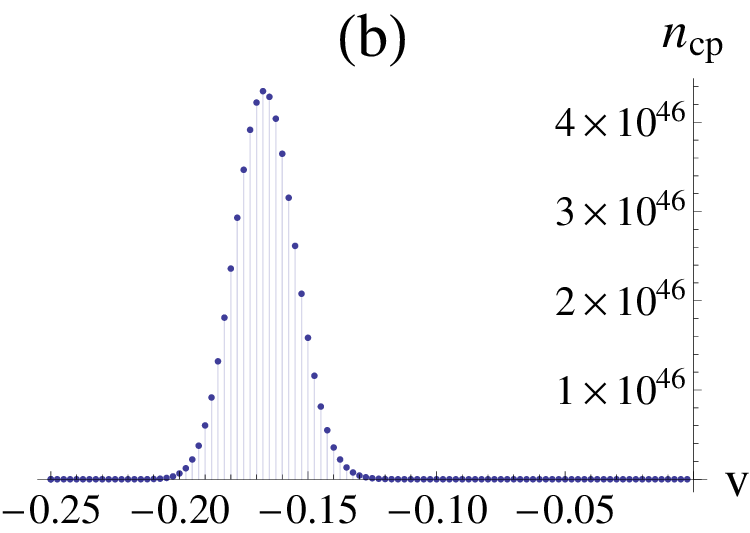}
\caption{Model (\ref{Vtradmf}) without interaction for $N=100$. (a) Logarithmic sum of critical points starting from left versus potential density. (b) Number of the critical points with respect to their critical potential density.}
\label{phi4noint_cp}
\end{center}
\end{figure}

\begin{figure}
	\begin{center}
		\includegraphics[width=0.35\textwidth]{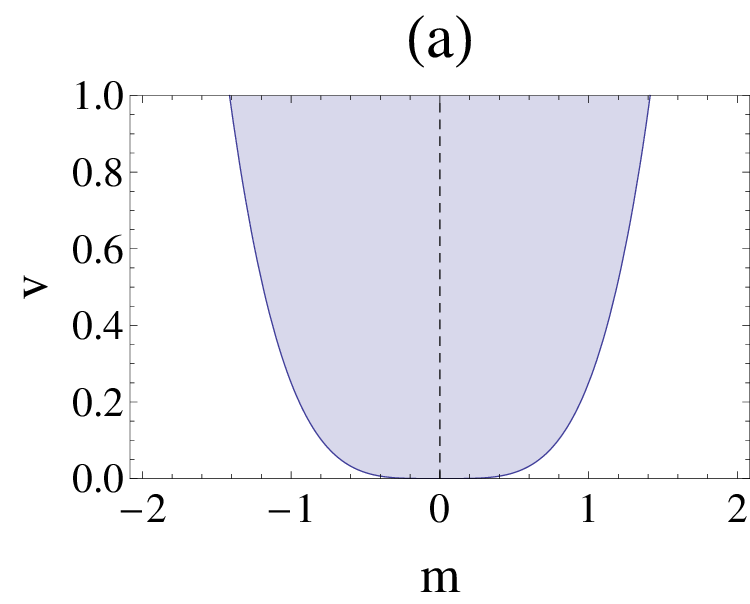}
		\includegraphics[width=0.35\textwidth]{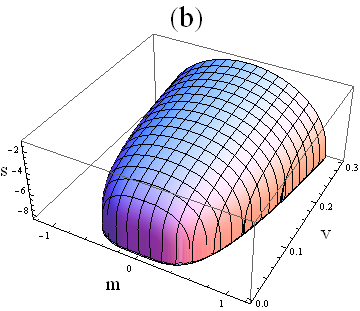}
		\caption{As Fig. \ref{pphi4_sdomain_fig} for the model (\ref{Vnoint}).}
		\label{phi4noint_entropy}
	\end{center}
\end{figure}
For comparison with the model (\ref{Vnoint}), in Fig. \ref{phi4noint_cp} we show the critical points of the model (\ref{Vtradmf}) without interaction which was investigated in \cite{b0}. The total large number of critical points is $3^N$, which is entirely due to the presence of the negative quadratic term in the local potential. $3^N$ comes from the combinatorial of the three solutions of the third degree equations inside the system $\nabla V=0$.

In Fig. \ref{phi4noint_entropy} we report the graphic of the entropy which is strictly concave, as it is the case for a system without SBPT.

\section{Short-range case}
\label{short}

Here, we conjecture that the models of the class (\ref{phi4model}) without a quadratic in the local potential still have a phase transition of the same type. We have no general demonstration available, but we provide some evidences that make this conjecture very reasonable. 

In this section, we will investigate the nearest-neighbor-interactions version of the mean-field simplified $\phi^4$ model introduced in Sec. \ref{pphi4mf}. The Hamiltonian is the following
\begin{equation}
V=\frac{1}{4}\sum_{i=1}^N \phi_i^4-J\sum_{\left\langle i,j\right\rangle} \phi_i\phi_j,
\label{Vsr}
\end{equation}
where we assume toroidal boundary conditions and $J>0$. The lattice sites can have any dimension $d$.

\subsection{Canonical thermodynamic}

By analysis, we can compute only the value of the spontaneous magnetization and of the specific potential at $T=0$. By inserting $q_i=q_0$ for $i=1,\cdots,N$ in (\ref{Vsr}) and dividing by $N$, we obtain 
\begin{equation}
	v=\frac{1}{4}q_0^4-dJq_0^2,
\end{equation}
from which, by vanishing the derivative
\begin{equation}
	\frac{\partial v}{\partial q_0} =q_0^3-2dJq_0=0,
\end{equation}
we get the solution for the spontaneous magnetization $q_0=0, \pm\sqrt{2dJ}$ and for the the specific potential $v_{min}=-d^2J^2$. $q_0=0$ has to be excluded for obvious reasons.

\begin{figure}
	\begin{center}
		\includegraphics[width=0.35\textwidth]{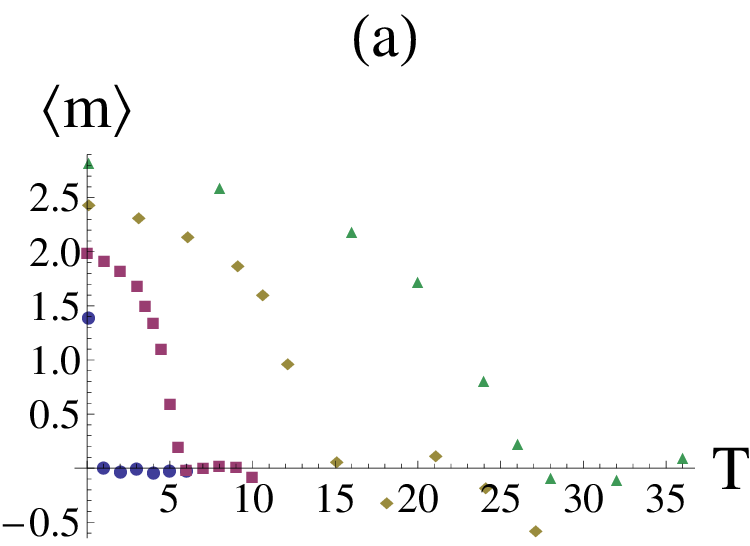}
		\includegraphics[width=0.35\textwidth]{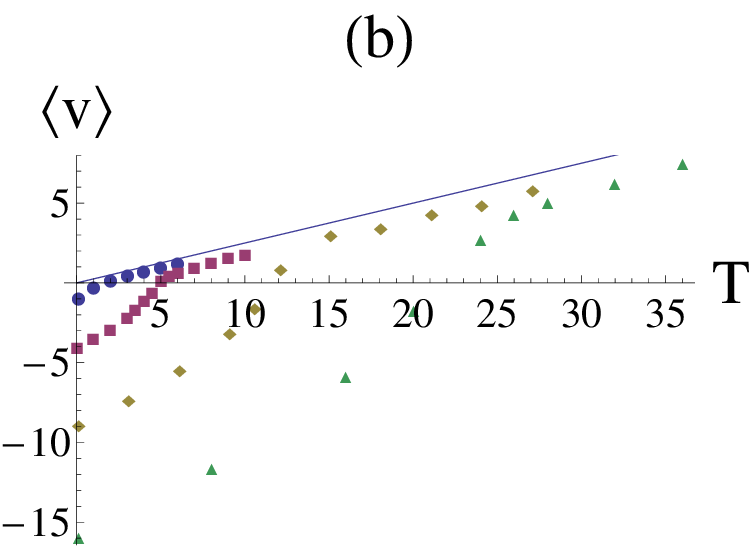}
		\includegraphics[width=0.35\textwidth]{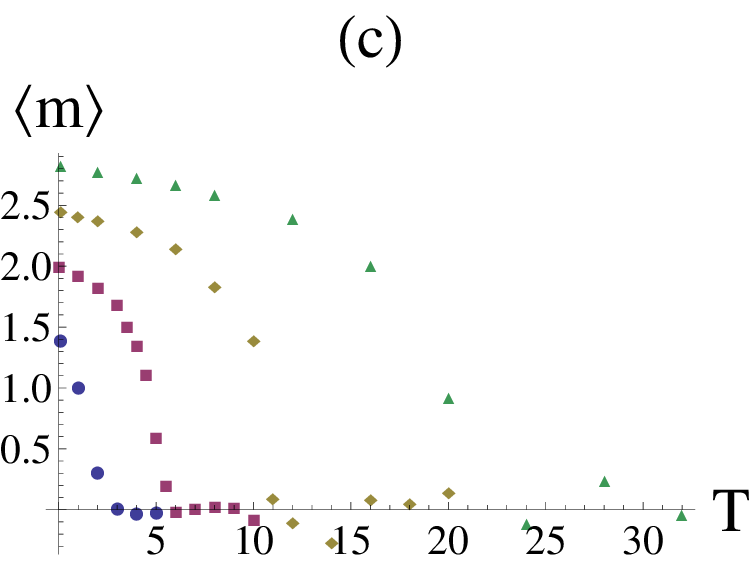}
		\includegraphics[width=0.35\textwidth]{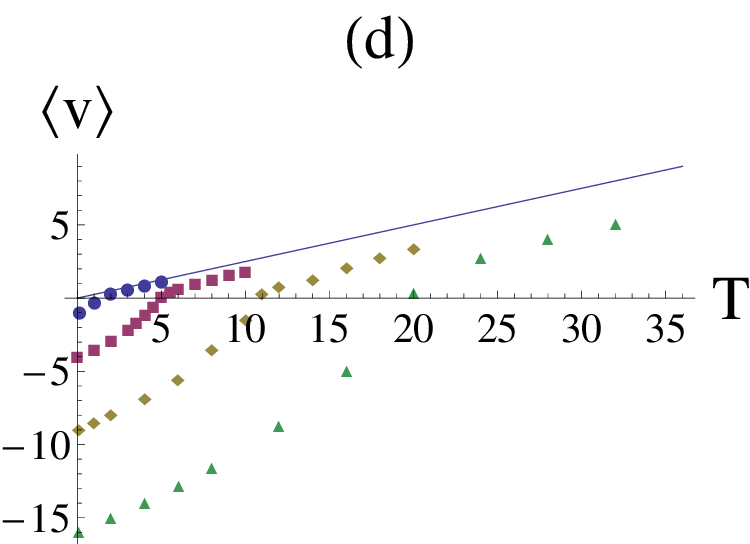}
		\caption{(a) Monte Carlo simulations for the spontaneous magnetization as a function of the temperature of the model (\ref{Vsr}) for $J=1$. Disks are for $d=1$ and $N=100$, squares are for $d=2$ and a 10 x 10 lattice, rhombuses are for $d=3$ and a 4 x 4 x 4 lattice, and triangles are for $d=4$ and a 3 x 3 x 3 x 3 lattice. (b) The same of (a) for the specific potential. The continuous line, $T/4$, is for the model without interaction. (c) The same of (a) for $d=2$ and a 10 x 10 lattice with $J=0.5, 1, 1.5, 2$ for disks, squares, rhombuses, and triangles, respectively. (d) The same of (c) for the specific potential.}
		\label{mTvTshort2d}
	\end{center}
\end{figure}

Some Monte Carlo simulations were carried out while varying $d$, and $J$. For $J=1$ the lattice dimensions are $d=1, 2, 3, 4$ with nearest-neighbors interaction, periodic boundary conditions for $d=1$ and toroidal for the other $d$-values. For $d=2$ the coupling constant is $J=0.5, 1, 1.5, 2$. The results are showed in Fig. \ref{mTvTshort2d}. As expected, the model shows a second-order phase transition except the case $d=1$ where the phase transition occurs at $T=0$. In general, we conjecture that the model (\ref{Vsr}) undergoes a $\mathbb{Z}_2$-SBPT for any $d$ and for any $J>0$ and belongs to the universality class of the classical Ising model in $d$ dimension, even extending the range of the interaction not only at nearest-neighbors. Among the $\mathbb{Z}_2$-SBPTs, we include also the special case $d=1$ as a limiting case with $T_c=0$.

Despite the difficulty of estimating precisely the critical temperature $T_c$ and the critical potential $\langle v\rangle_c$ as functions of $d$ and $J$ by those simulations, a quadratic relationship seems very reasonable. The rise of the minimum barrier between the two wells of the potential by increasing $d$ and $J$ explains the increase of $T_c$ and $\langle v\rangle_c$. The simulations suggest $\langle v\rangle_c>0$ for any value of $d$ and $J$.

\subsection{Critical points and topology of the equipotential hypersurfaces}
\label{srcptop}

$\nabla V=0$ for the potential (\ref{Vsr}) takes the form
\begin{equation}
	\phi_i^3-J\sum_{\left\langle i,j\right\rangle}\phi_j=0 \quad i=1,\cdots,N.
	\label{gradsr}
\end{equation}
A remarkable property for any lattice dimension $d$ is the existence of a scaling law that links the critical points evaluated at different values of the coupling $J$. The scaling is as follows
\begin{equation}
	\begin{cases}
		J\rightarrow J'
		\\
		\phi_i\rightarrow\phi_i'=k\phi_i \quad i=1,\cdots,N
		\\
		m\rightarrow m'=km
		\\
		v\rightarrow v'=k^4 v
	\end{cases},
	\label{scaling}
\end{equation}
where $k=\sqrt{J'/J}$. To prove it, consider
\begin{equation}
	k^3\left(\phi^3_i-Jk^{-2}\sum_{\left\langle i,j \right\rangle}\phi_j\right)=0\quad\quad i=1,\cdots,N.
	\label{}
\end{equation}
It is immediate to verify that if the set of coordinated $\phi_i$, $i=1,\dots,N$, is a critical point, then the new set $k\phi_i$, $i=1,\dots,N$, is a solution of $\nabla V=0$ for $J'$. The scaling holds also for the mean-field case. 

In \cite{b0}, it was analytically proven that all the critical levels of the model (\ref{Vtradmf}) are below zero. Here, we extend the demonstration to the model (\ref{Vsr}). From (\ref{gradsr}) we deduce that if $\mathbf{\phi}^s$ is a stationary point, then
\begin{equation}
	{\phi_i^s}^3=J\sum_{\left\langle i,j\right\rangle}\phi_j^s \quad i=1,\cdots,N.
	\label{gradsr_stat}
\end{equation}
Rewriting (\ref{Vsr}) in the following form
\begin{equation}
	V=\sum_{i=1}^N \phi_i\left(\frac{1}{4}\phi_i^3-J\sum_{\left\langle i,j\right\rangle} \phi_j\right),
	\label{Vsr_stat}
\end{equation}
and by substituting (\ref{gradsr_stat}), we get
\begin{equation}
	V(\mathbf{\phi}^s)=-\frac{3}{4}\sum_{i=1}^N {\phi_i^s}^4\le 0.
	\label{}
\end{equation}

In order to deepen a relationship with critical points and thermodynamic, let us define the minimum barrier $B_{min,N}$ between the two global minima of the potential at fixed $N$. To do this, let $p$ be a path, i.e. a continuous line, in configuration space which links the two global minima. For each point of $p$ the potential takes a value. Since the length of any $p$ is finite, then the set of the differences between the potential along $p$ and the global minimum $v_{min,N}$ of the potential has a maximum that we will call $V_p$. Then, we define $B_{min,N}$ as the infimum of the set of the $V_p$'s associated to all the possible paths $p$'s. We define also $b_{min,N}=B_{min,N}/N$.

There is an interesting relation between $b_{min,N}$ and the thermodynamic critical potential $v_c$. Let us assume that $\lim_{N\rightarrow\infty}b_{min,N}=b_{min}$ exists finite. Then, $v_c\geq b_{min}+v_{min}$ holds. This inequality is a consequence of Theorem $1$ in \cite{bc}. Indeed, for $v\in (v_{min}, b_{min}+v_{min})$ the $\Sigma_{v,N}$'s are topological equivalent to the disjointed union of two $N$-spheres, so that the hypotheses of Theorem $1$ are satisfied. The implication of the theorem is that the $\mathbb{Z}_2$ symmetry of the potential is broken for $\left\langle v\right\rangle\in [v_{min}, b_{min}+v_{min})$, whence $v_c\geq b_{min}+v_{min}$.

What can we say about $b_{min}$ as a function of $d$? Let us start with $d=1$. Suppose the configuration of the system is that of a global minimum, for example, $\phi_i=\sqrt{2J}$ for $i=1,\cdots,N$. Flip a degree of freedom, for example, $\phi_1=-\sqrt{2J}$ for fixing the ideas. The potential of the new configuration has increased by the quantity $4J$. Now, continue to flip the nearest neighbors until only one takes the value $\sqrt{2J}$. In so doing the potential does not change value. At this point we flip the last degree of freedom and the potential will return to its global minimum. We have so described a path in configuration space between the global minima of the potential for which $V_p=4J$. Since for $N\rightarrow\infty$ the above-defined path is that with the minimum $V_p$, then the minimum barrier is $B_{min,N}=4J$. 

For $d=2$ we can proceed in a similar way as follows. Consider a square lattice with toroidal boundary conditions. Flip the degrees of freedom belonging to a row, then flip the nearest neighbors until all the degrees of freedom are flipped. We have chosen a row (or a column is the same) because it is a path in configuration space of minimum length. Since the lattice is square, then the path length is $N^{1/2}$, so that $B_{min,N}\propto N^{1/2}$.

For any $d$, by the same considerations, we can show that $B_{min,N}\propto N^{(d-1)/d}$. The limiting case $d=\infty$ corresponds to the mean-field case for which we already showed in Sec. \ref{pphi4mf_geom} that $B_{min,N}\propto N$.

What could the implications of the minimum barrier be on the critical points? The $\Sigma_{v,N}$'s are topologically equivalent to two disjointed $N$-spheres in the $v$-interval comprised between $v_{min,N}$ and the first critical $v$-level. Assuming $b_{min,N}$ is the value of the minimum barrier implies that the $\Sigma_{v,N}$ corresponding to the potential $v_{min,N}+b_{min,N}$ is critical. If $b_{min,N}\to 0$ in the thermodynamic limit, at least a critical point exists whose critical $v$-value tends to $v_{min}$. This means that the interval $(v_{min,N},0)$ cannot be void of critical points. In Sec \ref{nphc}, we see that the considerations above are compatible with the results obtained using the NPHC method. 

It is worth noticing that this is strictly related to the convexity properties of the graphic of the entropy in the $(m,v)$-plane. The fact that $b_{min,N}\to 0$ in the thermodynamic limit makes it possible to divide the system configuration into at least two domains with independently oriented magnetization. This implies that all the magnetization values comprised between the maximum and the minimum are allowed for a fixed temperature. This reflects in a flat entropy as a function of $m$ at fixed $v$ comprised between the maximum and the minimum fo the spontaneous magnetization. In other words, the graph is non-strictly concave \cite{lanford, galla,hk,k2}. 

\begin{figure}
	\begin{center}
		\includegraphics[width=0.35\textwidth]{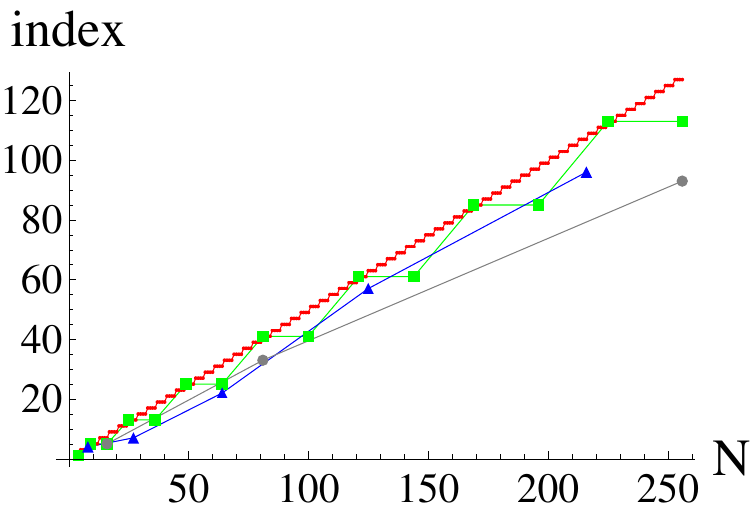}
		\caption{Index of the saddle point $(0,\cdots,0)$ of the model (\ref{Vsr}) as a function of $N$ for $d=1, 2,3, 4$ (red points, green squares, blue triangles, gray disks, respectively). The continuous lines are guides for the eye.}
		\label{pphi4Nindex}
	\end{center}
\end{figure}

It is possible to compute analytically via a computer algebra system the index of the saddle point $(0,\cdots,0)$, which grows linearly with $N$ at least for the values investigated here (see Fig. \ref{pphi4Nindex}). In the mean-filed case the index of the central saddle is $1$ independently on $N$. This is consistent with the fact the limit for $N\to\infty$ of the minimum barrier $b_{min}$ is finite in the mean-field case and is vanishing for finite $d$.

\subsubsection{Critical points using the NPHC method}
\label{nphc}

\begin{figure}
	\begin{center}
		\includegraphics[width=0.35\textwidth]{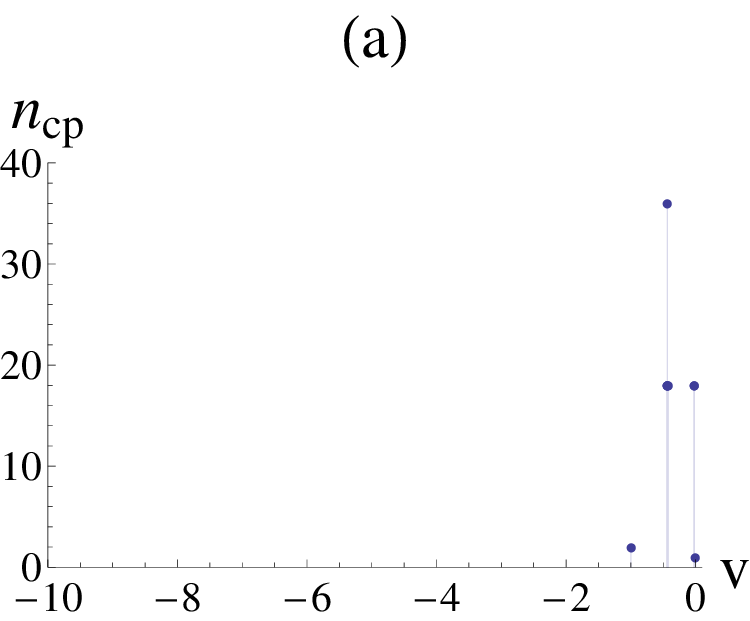}
		\includegraphics[width=0.35\textwidth]{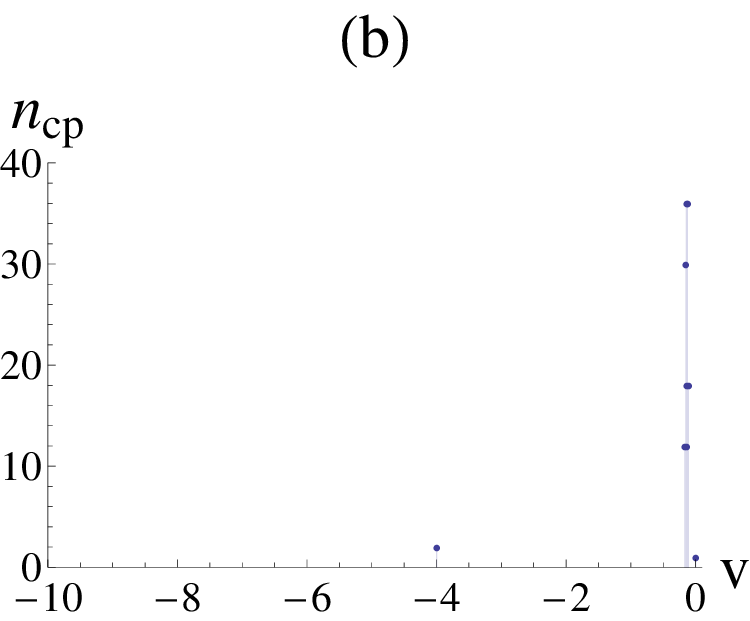}
		\includegraphics[width=0.35\textwidth]{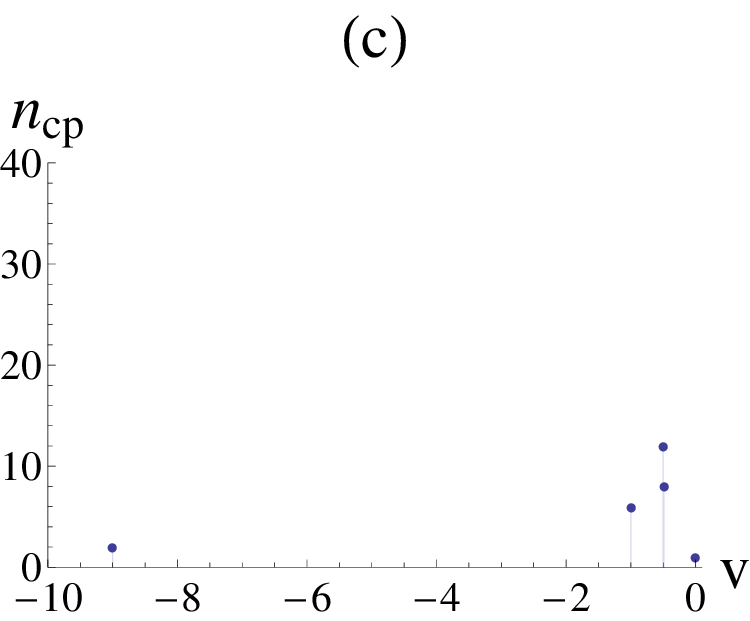}
		\includegraphics[width=0.35\textwidth]{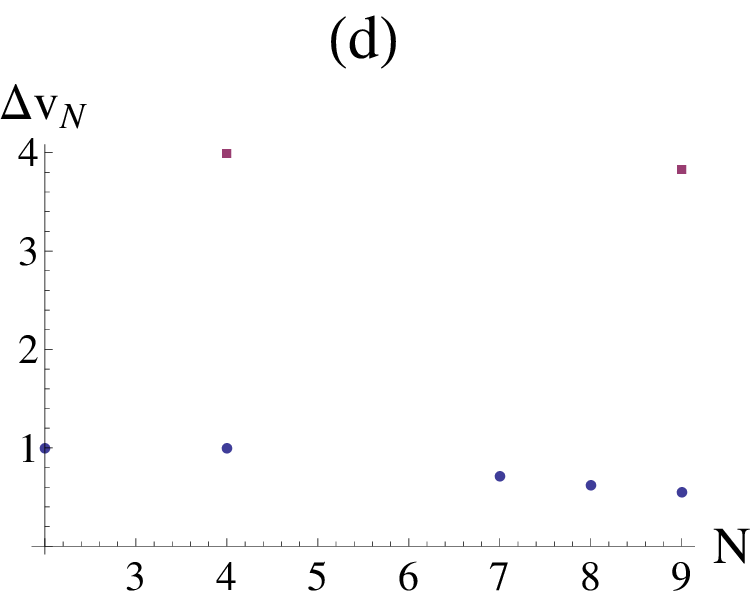}
		\caption{Nearest-neighbors model (\ref{Vsr}). Number of critical points founded using the NPHC method with respect to their critical potential density for $d=1$ and $N=9$ (a), for $d=2$ and $N=3 \times 3$ (b), and for $d=3$ and $N= 2\times 2\times 2$ (c). (d) Difference between the minimum critical potential level above and the global minimum for $J=1$ versus $N$ for $d=1$ (disks) and $d=2$ (squares).}
		\label{cpsr}
	\end{center}
\end{figure}

We used the NPHC (Numerical Polynomial Homotopy Continuation) method \cite{m} to solve (\ref{gradsr}) at least for values of $N$ up to $9$ and $d$ up to $3$. We chose the following cases to compute:
\begin{center}
	\begin{tabular}{l c}
		d=1  \quad N=4, 7, 8, 9\\ 
		d=2  \quad N=2x2, 3x3 \\ 
		d=3  \quad N=2x2x2 
	\end{tabular}.
\end{center}
\smallskip
The boundary conditions are periodic for the cases with $d=1$ and are toroidal for the other cases. The coupling $J$ is set to $1$ because the solutions of (\ref{gradsr}) for any other $J'$ can be obtained using the scaling law (\ref{scaling}). The results are reported in Fig. \ref{cpsr}. 

For $d=1$, we found an increasing total number of critical points while increasing $N$. In particular, at least up to $N=4$ the total number is $3$, which are the two global minima and the central saddle. For $N=7,8,9$ the total number is $31,67,147$, respectively. The smallness of the computed $N$-values did not allow us to make a significant statistic, but the data seems compatible with an exponential growth as in the case of the model (\ref{Vtradmf}). 

The most interesting feature is that the nearest critical $v$-level above the global minimum slightly lowers while increasing $N$ (see Fig. \ref{cpsr} panel (d)). This is a necessary condition if the minimum barrier $B_{min,N}$ for $d=1$ is independent on $N$, as it was shown in Sec. \ref{srcptop} for an $1$-dimensional lattice. For a consequence, $b_{min,N}$ decreases as $1/N$.

For $d=2$, the total number of critical points is $3$ in the lattice $2\times 2$ and some other critical points appear in the lattice $3\times 3$ with total number $165$. For $d=3$ some other critical points, beside the central saddle and the global minima, appear just in the lattice $2\times 2\times 2$ with total number $27$. 

Unfortunately, the computational means at our disposal did not allow us to investigate larger $N$-values. However, the data found confirm the hypothesis that the shape of the potential cannot be reduced to that of a double well with three critical points as for the model (\ref{Vmf}), expect in the case mentioned above for very small $N$-values and for $d=1,2$.


\section{Concluding remarks}

In this paper we show how a vanishing quadratic term of the local potential of the on-lattice mean-field $\phi^4$ model (\ref{Vmf}) with a $\mathbb{Z}_2$ ($O(1)$) symmetry entails a tragic simplification of the structure of the potential energy landscape. In particular, the number of critical points decreases to three. This has no influence on the $\mathbb{Z}_2$-SBPT properties, so that it makes easier the general study of the link between geometry and topology of the potential landscape and the phase transition. The only two critical levels of the potential are located at zero and at the global minimum. The critical potential of the $\mathbb{Z}_2$-SBPT is located above zero for any value of the coupling constant $J$. This clearly shows that the $\mathbb{Z}_2$-SBPT is not directly related to the presence of critical points, but, raver, to a particular shape of the $\Sigma_{v,N}$'s which can been defined 'dumbbell-shaped', which in turn are entailed by the presence of a double well in the potential with minimum barrier proportional to $N$. The last property explains the irreducible presence of three critical points because they are the minimum number requested for the occurrence of a double well in an analytic potential. 

The short-range case was also investigated. In this case the number of critical points cannot be reduced to three and no topology drastic simplification of the $\Sigma_{v,N}$'s can occur. For $N$-values up to $9$ and up to $d=3$ the presence of critical points in addition to the two global minima and the central saddle point $(0,\cdots,0)$ was detected using the NPHC method. Incrementing $J$ at fixed $N$ cannot cause any reduction of the number of critical points because a scaling law entails that their number does not change while varying $J$. The presence of critical points in addition to the global minima and the central saddle can be inferred also using thermodynamic reasons. In particular, the short-range interaction causes a minimum potential barrier between the two wells decreasing as $N^{-1/2}$, so that the $v$-interval between the global minimum and $0$ cannot be void of critical points. This means that the EP's cannot be homeomorphic to two disjointed $N$-spheres above the global minimum except for a $v$-interval that tends to $0$ in the thermodynamic limit.

A future plan is to extend this research to other symmetry groups.

\begin{acknowledgments}
I warmly thank Dhagash Metha for having suggested the application of the NPHC (Numerical Polynomial Homotopy Continuation) method to this study  and carried out the numerical computations needed to detect the critical points of the short-range cases of the $\phi^4$ model included in Sec. \ref{nphc}.
\end{acknowledgments}

\end{document}